# Challenges and Possible Strategies to Address Them in Rare Disease Drug Development: A Statistical Perspective

Jie Chen[1], Lei Nie[2], Shiowjen Lee[2], Haitao Chu[3], Haijun Tian[4], Yan Wang[2], Weili He[5], Thomas Jemielita[6], Susan Gruber[7], Yang Song[8], Roy Tamura[9], Lu Tian[10], Yihua Zhao[11], Yong Chen[12], Mark van der Laan[13], Hana Lee[2]*

[1]ECR Global, Shanghai, China

[2]US Food and Drug Administration, Silver Spring, MD, USA

[3]Pfizer Inc, Groton, CT, USA

[4]Eli Lilly & Co., Indianapolis, IN, USA

[5]AbbVie, North Chicago, Illinois, USA,

[6]Merck & Co., Inc., Rahway, NJ, USA

[7]TL Revolution, Cambridge, MA, USA

[8]Vertex Pharmaceuticals, Boston, MA, USA

[9]University of South Florida, Tampa, FL, USA

[10]Stanford University, Stanford, CA, USA

[11]Flatiron Health, San Francisco, CA, USA

[12]University of Pennsylvania, Philadelphia, PA, USA

[13]University of California at Berkeley, Berkeley, CA, USA

*Correspondence to: Dr. Hana Lee, US Food and Drug Administration, Silver Spring, MD 20903. Email: hana.lee@fda.hhs.gov.

# Challenges and Possible Strategies to Address Them in Rare Disease Drug Development: A Statistical Perspective

**Abstract**

Developing drugs for rare diseases presents unique challenges from a statistical perspective. These challenges may include slowly progressive diseases with unmet medical needs, poorly understood natural history, small population size, diversified phenotypes and geneotypes within a disorder, and lack of appropriate surrogate endpoints to measure clinical benefits. The Real-World Evidence (RWE) Scientific Working Group of the American Statistical Association Biopharmaceutical Section has assembled a research team to assess the landscape including challenges and possible strategies to address these challenges and the role of real-world data (RWD) and RWE in rare disease drug development. This paper first reviews the current regulations by regulatory agencies worldwide and then discusses in more details the challenges from a statistical perspective in the design, conduct, and analysis of rare disease clinical trials. After outlining an overall development pathway for rare disease drugs, corresponding strategies to address the aforementioned challenges are presented. Other considerations are also discussed for generating relevant evidence for regulatory decision-making on drugs for rare diseases. The accompanying paper discusses how RWD and RWE can be used to improve the efficiency of rare disease drug development.

# 1  Introduction

There is no universally acceptable definition for rare diseases. The World Health Organization (WHO) defines a rare disease as a health condition that affects less than 65 out of every 100,000 people [1]. The Orphan Drug Act (ODA) [2], a law passed by the United States (US) Congress in 1983 that incentivizes the development of drugs for rare diseases, defines a rare disease as a disease or condition that affects less than 200,000 (or 1 in 1,500) people in the US. The European Union defines a rare disease as a life-threatening or chronically debilitating disease with a prevalence rate less than 5 per 10,000 [3]. Recently China issued a report on rare diseases in which a rare disease is defined as a health condition that meets one of the following conditions (1) an incidence rate $< 1/10,000$ among infants, (2) a preva-lence rate < 1/10, 000, or (3) a total affected people < 140, 000 [4]. For regulatory purpose, rare diseases in Japan refer to as those with less than 50,000 affected individuals in order to be eligible for orphan drug designation in pharmaceutical regulation, which translates to a prevalence rate of approximately 4/10, 000 [5]. See Abozaid et al. [6] for a systematic review of criteria of defining rare diseases and orphan drugs in the regulatory setting.

Although individually rare, collectively over 10,000 rare disorders [7] are affecting approximately 400–475 million people worldwide [8], about two-thirds of whom begin in childhood [9]. Rare diseases are difficult to diagnose—an average time of 4-5 years for accurate diagnosis [10]—and even after an accurate diagnosis, treatment is often not available. Over the past 40 years after the ODA, the US Food and Drug Administration (FDA) has approved drugs with orphan drug designations covering only about 5% of rare diseases [11] and the majority of FDA-approved orphan drugs treat only one rare disease and have no other use [12].

Developing a treatment for a rare disease presents unique and arduous challenges, mostly due to the small number of individuals affected, early disease onset (often in infancy or early childhood), lack of accurate diagnosis, less understanding of disease natural history, and phenotypic diversity within a disorder. See Augustine et al. [13], Cheng and Xie [14], Kempf et al. [15], Fonseca et al. [16], Tambuyzer et al. [17], The Lancet Global Health [18] and also Section 3 for detailed discussion. Because of these challenges, there exist huge unmet medical needs for patients with rare diseases. For example, treatments for the 5% of rare diseases with authorized drugs are not necessarily transformative, leaving patients with only partially stabilized diseases and often seeking alternative options [19, 20]. Patients with the other 95% of rare diseases that have no authorized treatments available take their toll on all involved, many of them end up with premature death at infancy and young childhood or live with devastating long-term functional, physical, and/or mental disabilities [21, 22].

Although there exist substantial challenges in developing drugs for rare diseases, the evidence standard for regulatory approval of drugs for rare diseases remain unchanged. The FDA released in 2019 a draft guidance on "Demonstrating Substantial Evidence of Effectiveness for Human Drug and Biological Products" [23], which is complementary to the 1998 FDA guidance entitled "Providing Clinical Evidence of Effectiveness for Human Drug and Biological Products" [24]. The 2019 guidance further elaborates (1) accelerated approval pathway based on a demonstrated effect on a surrogate endpoint that is reasonably likely to predict a clinical benefit but where there are not sufficient data to show that it is a validated surrogate endpoint, and (2) approval based on one adequate and well-controlled trial plus confirmatory evidence. The latter includes, among others, certain types of real-world evidence (RWE). The agency acknowledges that certain aspects of drug development that are suitable for common diseases may not be feasible for rare diseases and that development challenges may be greater with increasing rarity of the disease. With all of these in mind, the Real-World Evidence (RWE) Scientific Working Group (SWG) of the American Statistical Association (ASA) Biopharmaceutical Section (BIOP), under the auspice of the Public Private Partnership (PPP) of the Center for Drug Evaluation and Research (CDER) in the FDA, has spent a significant amount of effort to evaluate the current landscape and possible use of real-world data (RWD) and RWE in drug development and regulatory decision for rare diseases. This paper is intended to provide a summary of landscape assessment including relevant regulatory guidance, current practice, challenges and strategies to address these challenges and the accompanying paper by the ASA BIOP RWE SWG [25] provides detailed descriptions on the use of RWD and RWE in rare disease drug development.

The rest of the paper is organized as follows. Section 2 summarizes regulations and guidance in drug development for rare diseases by major regulatory agencies. Section 3 discusses challenges in rare disease drug development from the perspectives of study design, conduct, regulatory approval, and other aspects. Strategies to address these challenges are presented in Section 4. Some concluding remarks are given in Section 5.

## 2  Regulatory Guidance and Related Documents for Rare Diseases

Regulatory agencies worldwide have introduced their respective guidance documents and frameworks covering a wide range of different aspects of drug development and regulatory decisions for rare diseases. The US FDA has issued perhaps more guidance documents for rare disease drug development than any other single agency in the world. We will therefore first summarize in this section relevant guidance documents by the US FDA and then by the other regulatory agencies.

## 2.1 Guidance by the FDA

In the past five years or so, the FDA has published the following guidance documents specifically for rare disease drug development. They are summarized as follows in the chronological order of their release dates.

- *Pediatric Rare Diseases–A Collaborative Approach for Drug Development Using Gaucher Disease as a Model* [26]. This guidance facilitates drug development in pediatric rare diseases and in particular, discusses a new possible approach to enhance the efficiency of drug development in pediatric rare diseases using Gaucher disease as an example.

- *Rare Diseases: Early Drug Development and the Role of Pre-IND Meetings* [27]. This draft guidance is intended to assist sponsors involving in drug development for rare diseases in planning and conducting more efficient and productive pre-investigational new drug (pre-IND) application meetings with the agency in dealing with challenges related to the nature of rare diseases.

- *Slowly Progressive, Low-Prevalence Rare Diseases with Substrate Deposition that Result from Single Enzyme Defects: Providing Evidence of Effectiveness for Replacement or Corrective Therapies* [28]. This guidance describes evidence necessary to demonstrate the effectiveness of INDs or new drug uses intended for slowly progressive, low-prevalence rare diseases that are associated with substrate deposition and are caused by single enzyme defects.

- *Rare Diseases: Common Issues in Drug Development* [29]. This guidance discusses issues that are commonly encountered and difficult to address in the context of rare disease drug development for which medical and scientific knowledge, natural history data, and drug development experience are often limited.

- *Rare Diseases: Natural History Studies for Drug Development* [30]. This draft guidance is intended to help inform the design and implementation of natural history studies that can be used to support product development for rare diseases. A natural history study collects information about the natural course of a disease in the absence of an intervention, from the disease's onset till either its resolution or death of the individuals. Although knowledge of a disease's natural history can benefit drug development for many disorders and conditions, natural history information may not be available or may be incomplete for most rare diseases.

- *Rare Pediatric Disease Priority Review Vouchers* [31]. This guidance provides, in the form of questions and answers, information on the implementation of priority review vouchers that allow the recipient to request the expedited review of certain rare pediatric disease product applications including definitions, policies, and procedures regarding rare pediatric disease product applications, requesting rare pediatric disease designation and priority review voucher, and relationship between rare pediatric disease designation and orphan drug designation.

- *Human Gene Therapy for Rare Diseases* [32]. This guidance provides recommendations to sponsors developing human gene therapy products for treating a rare disease in adult and/or pediatric patients regarding the manufacturing, preclinical, and clinical trial design issues for all phases of the clinical development program.

- *Rare Diseases: Considerations for the Development of Drugs and Biological Products* [33]. In addition to offering insights on natural history and nonclinical studies, this guidance also provides details on considerations for clinical pharmacology studies, dose selection, use of biomarkers, designs of clinical studies (e.g., innovative designs such as $n$-of1 design, randomized delayed-start designs, master protocols, and Bayesian methods), endpoint selection for effectiveness assessment, safety evaluation, and overall plans to maximize the quantity and quality of safety and efficacy data (including decentralized trials, comparator arm and expanded access program). The guidance also describes strategies for rare disease drug development in pediatric populations.

In addition, the FDA has recently issued guidance relating to clinical trial considerations to support accelerated approval of oncology therapeutics [34], the design and conduct of externally controlled trials [35], and the use of RWD and RWE to support regulatory decision-making [36]. Principles and recommendations presented in these guidance documents are also applicable to rare disease drug development.

## 2.2 Guidance and/or frameworks by other regulatory agencies

*The European Medicines Agency (EMA).* The EMA established in 2000 the Committee for Orphan Medicinal Products (COMP) which is responsible for evaluating applications for orphan drug designation [37]. To meet orphan designation, a medicine must be intended for the treatment, prevention, or diagnosis of a disease that is life-threatening or chronically debilitating with a prevalence of no more than 5 in 10,000 in the European Union (EU) and for which marketing of the medicine would not generate sufficient return to cover its development expenses.

The EMA published in 2006 a guideline on clinical trials for small populations which describes principles of clinical trials for rare diseases including pharmacological considerations, choice of endpoints, choice of control group, and design and analysis considerations [38]. In 2018, the EMA issued a question-and-answer document addressing common misunderstandings about orphan designation and related aspects pertaining to orphan medicines [39]. The EMA also issued in 2019 points-to-consider for the estimation and reporting of the prevalence of a condition for the purpose of orphan designation, which provides (1) some key definitions (e.g., prevalence, number of people affected, and population), (2) general considerations (e.g., geographic and temporal variation, chronological trend, point estimate of prevalence), and (3) prevalence of a condition (e.g., sources, validity, and comparability of data) [40]. Recently, the COMP of the EMA unveiled its pre-authorization activities aiming at (1) optimizing the quality of initial orphan designation applications, (2) ensuring consistency, transparency, quality, and detail of the grounds of opinions and orphan maintenance assessment reports, and (3) exploring cases and process options for RWE in orphan designation decision-making and the principles for future piloting of rapid RWE analytics to support regulatory decisions [41].

*The National Medical Products Administration (NMPA) of China.* A rare disease is defined in China as a disease with an incidence of less than 1/10,000 in newborns or a condition that affects fewer than 200,000 people in China [42]. The NMPA released in 2022 a technical guidance on clinical development for rare disease drugs which provides the sponsors with (1) special considerations in clinical development for rare disease drugs (e.g., obtaining clinical data of rare diseases, application of biomarkers and quantitative pharmacological tools, and establishment of patient registries), (2) clinical development plan, (3) design of clinical trials (e.g., precise definition of study population, choice of design, endpoint selection, sample size consideration), (4) requirements for safety evaluation, and (5) communication with authority [43]. The agency also published in 2022 the guidance on statistical considerations for drug clinical trials for rare diseases, in which clinical trial designs (e.g., response-adaptive design, *n*-of-1 design, seamless adaptive design, basket trials, Bayes methods, single-arm trials, real-world studies), sample size considerations, statistical analysis, special considerations during the conduct of clinical trials (e.g., selection of trial sites, special populations, eligibility criteria, recruitment, data collection, etc.), and assessment of evidence (e.g., efficacy and safety, surrogate and clinical endpoints, benefit and risk, etc.) [44]. More recently, the NMPA released a guidance on natural history studies of rare diseases for drug clinical development [45]. This guidance document provides (1) the application scope of natural history studies of rare diseases (e.g., identification of target population, assisting trial design, identification and validation of biomarkers, development and establishment of clinical outcome tools, serving as external controls, etc.), (2) types of natural history studies (e.g., retrospective, prospective, retro-prospective, and cross-sectional), and (3) implementation of natural history studies (e.g., planning and implementation stage, patient participation, data collection, and protection of patients).

*Ministry of Health, Labour and Welfare (MHLW) of Japan.* The Ministry of Health, Labour and Welfare (MHLW) of Japan established in 2009 an orphan drug/medical device designation system in which the designation criteria are described as that (1) the number of patients who may use the drug or medical device should be less than 50,000 in Japan, (2) the drugs or medical devices should be indicated for the treatment of serious diseases, including difficult-to-treat diseases, and (3) there should be a theoretical rationale for the use of the product for the target disease, and the development plan should be appropriate [46]. The MHLW recently issued a partial revision on the designation of orphan drugs that describes, among others, designation criteria including determination for the number of patients (criteria and method of estimation for the number of patients, and scope of diseases) and medical needs (seriousness of the target disease and effectiveness of the product) [47]. The revision is also explained in an companion questions and answers that further discuss some specific scenarios regarding designation of orphan drugs [48]. While the MHLW is responsible for review, pre-consultation, and final approval of orphan drug/medical device designation, the Pharmaceuticals and Medical Devices Agency (PMDA) of Japan provides priority scientific consultation for clinical trials and dossiers for marketing authorization of orphan drugs/medical devices. An overview on the application process for orphan drugs/medical devices designation can be found in MHLW [46].

*Canadian Agency for Drugs and Technologies in Health (CADTH).* The Canadian Organization for Rare Disorders (CORD) defines a rare disease as one that affects fewer than one in 2,000 people [49]. The CADTH released a framework on drugs for rare diseases with the objectives to launch the national strategy for the development, authorization, marketing, and incentives of drugs for rare disease [50]. The agency encourages sponsors to seek regulatory advice through pre-clinical trial application consultation meetings before their applications with respect to trial design, pediatric considerations, and regulatory requirements. In particular, the agency offers pathways for accelerating the regulatory review process for drugs for serious, life-threatening, or severely debilitating conditions for which there is substantial evidence for the drug to provide (1) effective treatment, prevention, or diagnosis of a disease or condition for which no drug is presently marketed or (2) a significant increase in efficacy and/or significant decrease in risk such that the overall benefit/risk profile is improved over existing agents [51].

*Ministry of Food and Drug Safety (MFDS) of South Korea.* The MFDS defines as a disease as one that affect no more than 20,000 people in South Korea [52]. The regulation on orphan drug designation by the agency requires sponsors to provide (1) data on the number of patients in Korea, (2) data proving to be used for a disease for which an appropriate drug has not been developed or that the drug has significantly improved safety or efficacy over existing products, and (3) a recommendation letter for orphan drug designation from relevant professional association. In addition to financial incentives and extension of expiration from 5 years to 10 years, the regulation also offers exemption from re-examination and data master file registration for orphan drug designation [53].

Since the United Nations (UN) adopted the first-ever UN Resolution on "Addressing the Challenges of Persons Living with a Rare Disease and their Families", rare diseases have been recognized by health organizations and individual countries/regions as one of the most important health problems to be addressed [54]. Accordingly, health professionals have called for actions to address these challenges and propose recommendations to policy makers that aim to encourage, among others, the development of drugs and treatments for people living with a rare disease, e.g., Giugliani et al. [55], Wainstock and Katz [56].

## 3 Challenges in Rare Disease Drug Development

Although many of the challenges faced in drug development for common diseases could also apply, there are special challenges related to rare disease drug development. This section delineates challenges and considerations from a statistical perspective in the design and conduct of clinical trials specifically for rare disease drug development.

### 3.1 Study design

*Small population size.* The limited number of patients creates challenges to recruiting adequate number of patients in a randomized controlled trial (RCT) to demonstrate clinically meaningful treatment benefits with a sufficient statistical power. For example, a study found that over one-fourth of rare disease clinical trials conducted between 2016 and 2020 terminated early due to low patient accrual rates [57]. In addition, results from a clinical trial with a small sample size, which is usually the case for rare disease trials, may not be generalizable to a broader patient population. See Augustine et al. [13], Cheng and Xie [14], Kempf et al. [15]; and [17] for detailed discussions.

*Lack of accurate diagnosis.* The diagnosis of rare diseases presents numerous challenges due to the rarity and often complex nature of these disorders: (1) many healthcare providers may not be familiar with rare diseases, which can lead to delayed or incorrect diagnosis or even misdiagnosis; (2) diagnosis of many rare diseases requires specialized tests or procedures that may not be readily available in some healthcare settings, which can result in delayed or incomplete diagnostic evaluation; (3) heterogeneous manifestations of some rare diseases present special challenges to establish definitive diagnosis; (4) many rare diseases are caused by unknown gene mutations, which makes it challenging and time-consuming to identify specific gene alteration that is responsible for the disorder; (5) other factors may include high cost of diagnosis, less financial incentives to development diagnostic tools, and lack of comprehensive research data on rare diseases. See Marwaha et al. [10], Austin et al. [58], Casas-Alba et al. [59], Zanello et al. [60] and Adachi et al. [61] for more discussions.

*Limited understanding of natural history.* The information of a disease's natural history is critical for planning drug development, e.g., identifying target patient population, defining appropriate endpoints, and developing biomarkers to guide clinical development strategies. The natural history of a disease, especially a rare disease, can also help determine external controls, either concurrent or non-concurrent [17, 30]. However, the natural history of disease is usually under-explored and hence inadequately characterized for most rare diseases [62, 63]. Many rare diseases have substantial genotypic and/or phenotypic heterogeneity, and the natural history of each subtype may be even more poorly understood, which intensifies the need for prospectively designed protocol-driven longitudinal natural history studies to generate more information on the natural course of a disease. This will in turn help evaluate whether a treatment changes the course of disease progression or affects the survival or quality of life of the patients [64, 65].

*Lack of consensus on clinically meaningful endpoints.* The choice of clinically meaningful and relevant endpoints that help measure treatment effects (or lack thereof) is fundamental to the design of clinical trials seeking regulatory approval. This requires understanding of the natural history of the disease without intervention or with standard of care [66]. For many rare diseases, lack of well-defined, practical, and clinically meaningful endpoints due to limited understanding of disease's pathophysiology may hamper regulatorily acceptable clinical trial designs. In addition, many slowly progressive rare diseases requires a long treatment duration, which makes it difficult to capture the disease natural history [67] and impractical to use clinically meaningful endpoints that require a long follow-up. Heterogeneous clinical presentations also make it hard to select a single primary endpoint that is clinically sensitive to all manifestations of the disease.

*Use of surrogate endpoints.* Clinical trials in rare diseases often use surrogate endpoints to measure therapeutic effects, due to the nature of the diseases (e.g., slowly progressive) and unmet medical needs, for regulatory accelerated approval. Such an approach has an obvious limitation of uncertainty in translating the surrogates into clinical benefits on how the patients feel, function, and survival [67–70].

*Long-term trials.* Many rare diseases are slowly progressive and thus require long-term (either single-arm or randomized controlled) studies to demonstrate clinical effectiveness of an investigational product. This may pose recruitment and retention challenges because patients may not be willing to participate or stay in a long-term trials.

## 3.2 Study conduct

The major obstacles during trial conduct can be summarized as follows.

*Trial perspective.* Clinical trials are often conducted in academic medical centers which may not easily be accessible to patients with rare diseases. Patients with many rare diseases are geographically dispersed, leading to the challenge of enrolling a sufficient number of patients in a trial that often requires participation of many sites across the US or sites from multiple countries (which may define a rare disease differently) [13, 15].

*Patient perspective.* Lack of awareness from patients about the rare disease may affect their attitude and willingness to participate in clinical trials. Lack of trust in the healthcare system and of supportive services (transportation, financial assistance, and language translation) may delay recruitment [71, 72].

*Community and healthcare system perspective.* There may be insufficient engagement of heathcare system and local communities in rare disease clinical trials due to, e.g., inadequate funding, resource, and comprehensive specialists, lack of awareness and trial-related information in local communities, and lack of prioritization and integration of rare disease trial options for patients (in case of multiple ongoing trials) [73–75].

*Other challenges.* Insufficient connection and partnership among patient organizations, physician networks, and medical societies and inflexibility of outcome assessment and its extensive documentation can delay enrollment. In addition, medical licenses for physicians issued by individual states may limit the ability of trial physicians to visit patients across state lines [76, 77].

## 3.3 Statistical analyses

The challenges for study designs discussed in Section 3.2 post unique difficulties for statistical analyses which are summarized below.

*Externally controlled trials (ECTs).* Many clinical trials for rare diseases are designed as externally controlled trials such as single-arm trials (SATs) and RCTs with internal control augmentation, for which statistical inference for treatment effects often relies on external data and evidence. This may lead to biased estimates of treatment effects due to patient selection and unmeasured confounding [78–80].

*Heterogeneity in patient population.* Many rare diseases are heterogeneous with respect to patient genotypic and phenotypic characteristics, some of which may be unknown or not be incorporated into trial designs and analyses. This can lead to higher-than-expected variability of the analysis variables and thus insignificant results [15, 81, 82].

*Surrogate endpoints, clinical benefits, and statistical power.* As discussed in Section 3.1, surrogate endpoints are often used in rare disease trials for earlier readouts but may not be translated into clinical benefits which are required for full approval of a medical product. To this end, it is important in the statistical analysis to consider the predictability of surrogates to clinical benefits [83, 84]. In addition, rare disease trials often enroll small sample sizes of patients which may not provide sufficient statistical power to produce statistically significant results on primary endpoints [85, 86].

*Benefit-risk assessment (BRA).* The BRA is quite complex in any clinical trials, requiring considerations on multiple aspects, e.g., analysis condition, alternate treatment options, benefits, risks, and risk management [87, 88]. It is even more challenging for rare disease trials due to, e.g., single-arm design (no comparison group for BRA) and short follow-ups (inadequate clinical benefits observed) [89].

## 3.4 Others challenges

Other challenges may include:

- *Lack of enough market incentive.* The ODA grants the sponsor of an orphan drug that is approved by the FDA the market exclusivity for 7 years (on the contrary, only 5 years for drugs treating a common disease) and a tax credit of up to 50% (cut to a half in 2017) for qualified clinical testing expenses [2, 90]. However, the current financial incentives are still insufficient to steer the drug development for rare diseases (especially for extremely rare diseases) due to smaller markets [13, 91–93]

- *Lack of high-quality RWD.* Rare diseases are often diverse and have multiple subtypes, having no standard approach to collect natural history data, which can lead to bias in selective enrollment, missing data, and other challenges in deriving high-quality data. In addition, lack of standard, clinical practice, and health policy for rare diseases across countries makes it difficult to integrate data from registries (or other sources) for rare diseases to generate more reliable information to support drug development.

- *Pediatric populations.* Most rare diseases ( 70% –75%) have pediatric onset. In general, pediatric drug development is driven by adult drug development when the target disease occurs in both groups, which causes an average of nine-year delay in pediatric drug approval after initial adult drug approval [94], even with the success of the Pediatric Research Equity Act (PREA), a huge gap in pediatric drug development still persists [95]. In addition, the usual development pathway from adults to pediatrics may not be viable for diseases (including rare diseases) that only affect pediatric populations, e.g., diffuse intrinsic pontine glioma (a rare malignant pediatric brain tumor) [96].

- *Lack of coordinated effort.* The current rare disease research is very siloed and often focused on single disorders, which creates challenges for joint effort, sharing data, assessing outcomes, and capturing knowledge that might be relevant across diseases [97, 98]. Halley et al. [99] point out that the single-disease approach limits our ability to (1) increase the efficiency of coordinated outcome development and test for rare diseases with similar phenotypes, etiology, or trajectory, (2) investigate the same biological pathways that are shared by several diseases, and (3) simultaneously assess the value of diagnostic tools across diverse rare disease communities. A recent survey by Clinical Research Networks (CRNs) for rare diseases finds that barriers preventing or slowing down rare disease drug development at the international front include lack of (1) sufficient dedicated public and private funding to support international projects on rare diseases, (2) harmonization in regulatory and contracting processes for data sharing, IRB reviews, national clinical trial guidance, and rules governing collaboration with industry and intellectual property rights, (3) common data standards and platforms, which hampers data sharing, (4) a governance framework between networks for coordination and management effort, and (5) an international network for global collaborative effort and actions [77, 100]

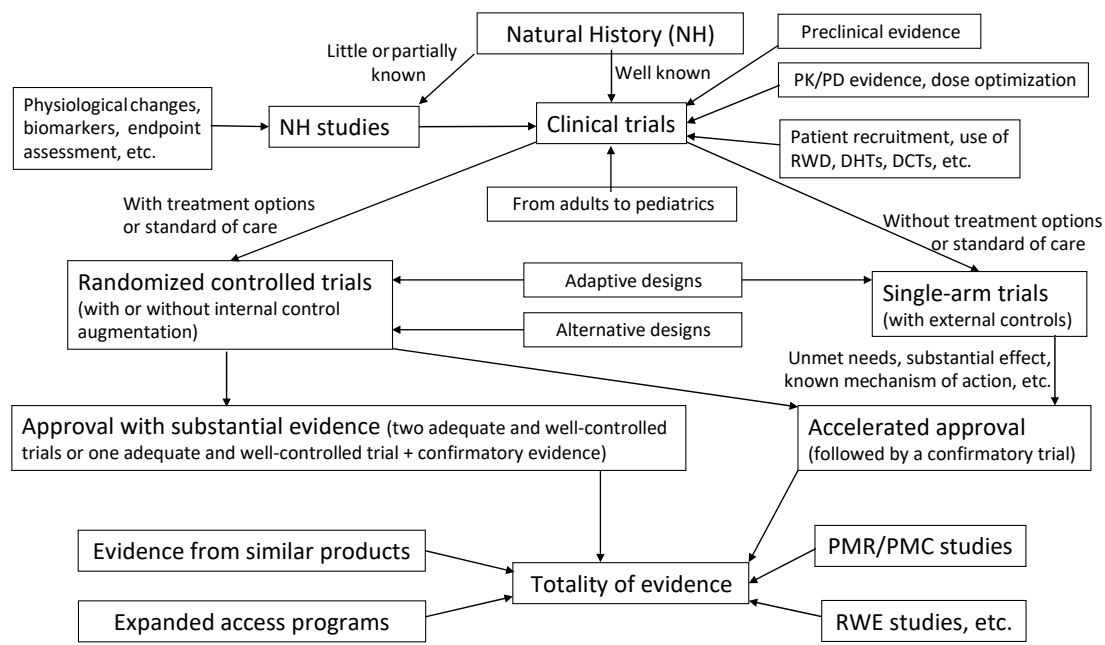

Figure 1: Pathways (including development pathways) for generation of totality of evidence for drugs treating rare diseases. DCT–decentralized clinical trials; DHT–digital healthcare technologies; MoA–mechanism of action; PD–pharmacodynamics; PK–pharmacokinetics; PMC–post-marketing commitment; PMR–post-marketing requirement; RWD–real-world data; RWE–real-world evidence.

# 4 Strategies to Address the Challenges in Rare Disease Drug Development

To address the aforementioned challenges, some strategies can be considered from the perspectives of trial design, conduct, and analyses. The rest of this section discusses possible strategies that can potentially be used to help address the challenges described in the previous section. Of note, before choosing appropriate strategies, it is important to understand the common evidence generation pathway (including development pathways) for drugs treating rare diseases, which can be illustrated in Figure 1.

## 4.1 An overall development strategy

Approvability of an IP is based on substantial evidence of effectiveness as defined in the Code of Federal Regulations (CFR) [101, 21CFR514.4] and Subpart E of the CFR on "Drugs Intended to Treat Life-Threatening and Severely Debilitating Illnesses" [101, 21CFR312.82] calls for the FDA to exercise broad scientific judgement in applying the evidential standards in decision on approval when no satisfactory alternative therapy is available. The FDA guidance on "Demonstrating Substantial Evidence of Effectiveness for Human Drug and Biological Products" [23] in which two adequate and well-controlled clinical trials or one adequate and well-controlled clinical trial plus confirmatory evidence are considered meeting the substantial evidence criteria. Regarding confirmatory evidence, the FDA guidance [102] provides examples on the following types of confirmatory evidence that can be used to substantiate one adequate and well-controlled clinical trial to demonstrate substantial evidence of effectiveness: (1) clinical evidence from related indication (e.g., evidence from a trial for a closely related indication), (2) mechanistic or pharmacodynamic evidence by fully understanding the pathophysiology of the disease and the mechanism of action of the drug, (3) evidence from preclinical studies (e.g., relevant animal models), (4) evidence from other drugs in the same pharmacological class approved for the same indication, (5) natural history evidence (e.g., outcomes observed can accurately reflect those that would have been expected in the absence of the treatment), (6) real-world data and evidence (e.g., one or a variety of sources of RWD to generate reliable RWE), and (7) evidence from expanded access use of an IP (e.g., the use of an IP for diagnosing, monitoring, or treating a disease other than in a clinical trial).

As noted in FDA [23], a single-arm trial with an external control may be an appropriate option for generating effectiveness of drugs when the natural history of a disease is well defined, the external control and treatment population are similar, concomitant treatments are not substantially different between the external control and treatment groups, and the results are compelling. From statistical perspective, treatments for rare diseases are often intended to address unmet medical needs and hence balancing the false positive and false negative results should be considered. In addition, statistical methods to assess therapeutic effects of drugs for rare diseases should consider the feasibility of trial design, sample size, and endpoints when demonstrating substantial evidence in these settings FDA [23]. When using single-arm trials to support accelerated approval of drugs for rare diseases, the principles discussed in FDA [34] for oncology drug development may also apply to rare diseases in general.

## 4.2 Trial design

*Adaptive designs.* Given the challenges in study design discussed in Section 3.1, adaptive designs [103] are perhaps one of the most suitable study designs for rare disease clinical trials. The FDA guidance on rare diseases [29] states that identifying early biomarkers of diseases or effects of interventions can be considered in adaptive and enrichment designs for greater efficiency. Kempf et al. [15] point out that prospectively planned adaptive designs may be used to combine dose finding trials with the larger late phase trials by screening and enrolling patients from the early phase trials into the late phase trials and dropping arms that lack favorable efficacy results and to enrich the patient population upon an interim analysis with improved understanding of responder characteristics. For example, the small sample sequential multiple assignment randomized trial (snSMART) design can be appropriate for rare diseases that are chronic, relative stable over the two stages of the trial in which all patients are randomized to a set of active (or some dose of) treatment options and may be re-randomized based on response to prior treatment [104]. Such a design could improve patient engagement and retention by allowing for patients to switch between treatments (doses) and study efficiency by combining both dose-finding and confirmation of therapeutic effects into one trial for product registration. In case of uncertainty about treatment effect on multiple endpoints that would be considered as primary endpoints for regulatory approval, the adaptive endpoint selection design can be considered to learn about disease endpoints through a subset of patients, in which the adaptation rule is pre-specified and type I error is controlled for adaptive endpoint selection, e.g., the use of combination test following the partition test principle in Xu et al. [105].

*Alternative designs.* For rare diseases or life-threatening or severely debilitating illness with an unmet medical need, the FDA recognizes the feasibility issues regarding certain aspects of drug development and the increasing development challenges with increasing rarity of the disease. To address the need, the FDA guidance on substantial evidence [23] recommends considering the advantages and disadvantages of various trial designs to achieve the objectives of establishing evidence of safety and effectiveness, including an RCT with unequal allocation (to ensure more patients receiving the IP than the control) or a single-arm trial with an external control. Sponsors may consider addressing recruitment and retention issues by using modified trial designs, including dose-response randomization, crossover trials, randomized withdrawal design, adaptive designs, or randomized delayed start design [29]. Both randomized withdrawal and delayed start designs are 2-stage designs; in the former, all patients receive open-label treatment during the first stage from which responders are randomized to treatment or placebo during the second stage, while in the later, patients are randomized to treatment or placebo in the first stage, followed by a second stage during which all patients receive active treatment [106, 107]. In addition, novel trial designs such as master protocols to include multiple IPs (e.g., a basket trial) or multiple subtypes of a rare disease (e.g., an umbrella trial) in a single protocol, with or without a common internal control, can be considered for flexibility and efficiency [108].

*Expansion from adult to pediatric population.* Drug development generally follows the usual pathway from adult programs to adolescent programs which in turn inform the pediatric programs, where the model-informed extrapolation can be used for trial design in younger population [109, 110]. However, different pathways may be considered, depending on whether the drug under development is for both adult and pediatric populations, or exclusively for a pediatric population. Kempf et al. [15] point out that careful planning for drug development of rare pediatric diseases is required to maximize the efficiency and increase the likelihood of success of clinical development.

*Use of biomarkers.* Biomarkers measure the existence and/or progression of diseases and therapeutic effects of the treatment and can be classified as diagnostic, pharmacodynamic or response, predictive, prognostic, safety, and risk biomarker [111]. Depending on the proposed use, all these biomarkers are closely related to rare disease drug development. For example, a diagnostic biomarker helps precise diagnosis of a rare disease, a response biomarker helps measure the effects of treatment, and prognostic biomarker help determine prognostic outcome of a rare disease. Knowledge about a disease's pathophysiology and clinical manifestations over time is frequently incomplete for rare diseases, which prompts further research in identification and validation of biomarkers that could be used in rare disease clinical trials (e.g., biomarker-driven adaptive design). For rare diseases that are serious or life-threatening with no efficacious therapies available, one may consider less established response biomarkers (surrogate endpoints that can reasonably likely predict clinical benefits) to measure drug effect as a basis for accelerated approval [29]. However, the use of a surrogate endpoint requires clinical and analytic validation of the biomarker [112].

*Choice of endpoint.* Selection of the most appropriate and sensitive efficacy endpoints is challenging in rare disease trials. The FDA established in October 2022 a Rare Disease Endpoint Advancement (RDEA) Pilot Program to promote innovation and evolving science in support of novel efficacy endpoint development for drugs treating rare diseases [113]. Under the RDEA Pilot Program, the FDA will give preference to proposals that (1) have broader potential impact on drug development, (2) reflect a range of different types of endpoints, and (3) use novel approaches for collection of additional clinical data in development stage in support of validation of surrogate endpoints. Of note, due to lack of consensus and limited historical data to inform the selection of one specific primary endpoint in many rare diseases, multiple competing endpoints may be considered in the trial design.

*Choice of external controls.* When preclinical and/or early clinical evidence shows a promising efficacy effect, an RCT with smaller internal concurrent control arm or a single-arm trial with an external control arm (when an internal control arm is impractical, e.g., no other treatment option) can be considered. ICH [114] describes a variety of controls including historical controls and the recently issued FDA guidance [35] provide thorough discussions and considerations for externally controlled trials with respect to the design, data sources for external controls, statistical analysis, and communication with the FDA.

## 4.3 Trial conduct

Multi-faceted strategies can be taken to address challenges in the conduct of rare disease trials.

*Use of RWD for patient accrual.* RWD databases such as disease registries for rare diseases can be used to provide the distribution of patients with different demographic factors (e.g., phenotyping features) and geographic locations (e.g., by zip code) that can help set patient eligibility criteria, choose trial sites, and enroll patient [115].

*Use of digital healthcare technologies.* Use of digital healthcare technologies (DHTs)

(e.g., telehealth, telemedicine, wearable devices) during trial conduct provides great opportunities to make a clinical research model inclusive, accessible, and efficient. DHTs are often used in DCTs to help enhance convenience for rare disease patients with limited mobility and improve their engagement, recruitment, and retention [116].

*Decentralized clinical trials (DCTs).* With the use of DHTs, rare disease trials can be performed by including some decentralized elements such that trial-related activities occur at locations other than traditional trial sites [117]. DCTs can be particularly beneficial to rare disease patients who are usually geographically dispersed and/or physically inconvenient. See Ghadessi et al. [118] for more discussions for the planning and conduct of DCTs with a special focus on rare diseases.

## 4.4 Statistical analyses

Methods of statistical analyses for rare disease trials should be based on the trial objectives and corresponding design; see Chow [119], Yang et al. [120] for more description on methods for statistical analyses for rare disease trials. This section discusses possible solutions to the challenges discussed in Section 3.3.

*SATs with external controls.* Methods for externally controlled trials such as Bayesian borrowing, propensity score and related methods, and *G* methods can be used to estimate treatment effects. Attention should be given to whether the patients in the SAT are comparable with those in the external controls with respect to their baseline (and possibly time-varying) covariates. Adequate sensitivity analyses should be performed to assess potential biases possibly caused by patient selection and unmeasured confounding [78].

*Multiple endpoints, global tests, and repeated measures.* In addition to composite endpoint approach (e.g., Verbeeck et al. [121]), methods such as win-ratio [122, and its variation] and desirability of outcome ranking [123, and its related methods] can be used to integrate evidence from multiple endpoints with different clinical interpretation and to address the challenges associated with different types of endpoints (e.g. clinical event, functional assessment, biomarker, and patient reported outcomes). For diseases that affect multiple organs and tissues and have heterogeneous clinical presentations, global tests for multiple endpoints that measure outcome from different organs may improve the study power. The tests may further provide a broad efficacy assessment for novel trials that use different endpoints for different patient subpopulations [124, 125]. In addition, when appropriate, longitudinal and repeated measurement data analyses can be advantageous to address the major concern of limited patients available for rare disease trials by reducing outcome variability [126].

*Predictability of surrogates for clinical benefits.* One may consider hierarchical models to investigate treatment effects using surrogate endpoints (first-layer model) which are then used to study overall survival (the second-layer model) [127–130]. The first-layer model can be used in early decision for regulatory consideration and the second-layer model serves as connection between the surrogates and overall survival.

*BRA.* Besides the general principles for a structured and well-defined BRA process as described in relevant guidance [88, 131], additional considerations may include: (1) some specific tools such as the magnitude of clinical benefit scale [132] can be used to derive reliable and fair evaluation of benefits, (2) multiple endpoints with ranking analyses such as the win-ratio and desirability of outcome ranking can be considered in the BRA, (3) for externally controlled trials, the BRA should take into account the clinical context (e.g., the consequence if patients do not receive the treatment under investigation) and the benefit-risk profiles of external controls. Overall, the BRA for a rare disease drug should reflect the rarity, seriousness, and possibly disease-induced disability or mortality of the condition, quality of life of patients, innovative nature of the therapy, and possibly perceived medical added-value to currently available treatments [133, 134].

## 4.5 Other considerations

*Post-marketing studies.* Accelerated approval of a rare disease drug using surrogate endpoints, often mandates post-marketing requirement studies to confirm clinical benefits of the drug [34, 135]. PMR studies for rare diseases may include observational studies, pragmatic trials, RCTs, or meta-analysis with clearly defined timetable for completion [136, 137]. Failure to comply with the timetable for completion of PMR milestones is patient to enforcement action in the absence of a demonstration of good cause [138].

*Expanded access program (EAP).* The EAP is a potential pathway for patients with rare diseases that are serious or life-threatening with no satisfactory treatment options to gain access the IP for treatment outside of clinical trials [139]. The EAP may use less stringent eligibility criteria (compared to a randomized controlled trial), a randomized withdrawal design, and/or open-label extensions for participation of patients who are likely exposed to the IP upon approval and to provide more information on dosing, safety, and efficacy [15, 140].

*Use of RWD and RWE.* The Framework for FDA's Real-World Evidence Program [141] outlines the scope of the FDA's RWE program and the plan to be used to implement it. The Framework specifically states that "RWD can help with...assembling geographically distributed research cohorts (e.g., in drug development for rare diseases..." and that "In limited instances, FDA has accepted RWE to support drug product approvals, primarily in the setting of oncology and rare diseases." The accompanying paper by the ASA BIOP RWE SWG discusses in more details about the scope, current practice and proposed use of RWD and RWE in drug development, a targeted-learning roadmap, and several case studies on the use of RWD and RWE in rare disease drug development and regulatory decision-making [25].

Given the above discussions, the challenges and corresponding strategies to address them can be summarized in Table 1.

Table 1: A summary of challenges and strategies to address them in drug development of rare diseases.

| Trial Aspects | Challenges | Strategies |
|---|---|---|
| Trial design | <ul><li>Small population size to prevent recruiting adequate number of patients</li><li>Lack of accurate diagnosis precluding identification of well-defined patient population</li><li>Limited understanding of natural history to hinder identifying target population, defining appropriate endpoints, and developing biomarkers</li><li>Lack of consensus on clinically meaningful endpoints to limit the selection of endpoints that are clinically sensitive to treatment</li><li>Use of surrogate endpoints that may not be translated into clinical benefits</li><li>Long-term studies to cause patient compliance issues</li></ul> | <ul><li>Adaptive design to accommodate evolving information and small sample size (e.g., snSMART) for efficient development program</li><li>Alternative designs to include RCTs with unequal allocations, dose-response randomization, crossover design, master protocol design, etc.</li><li>Expansion from adult to pediatric population pathway to tackle the needs of pediatric rare diseases</li><li>Use of biomarkers for disease diagnosis, patient identification, and treatment effect measurement</li><li>Appropriate choice of possibly multiple endpoints to capture various aspect of treatment effect</li><li>Use of external controls improve efficiency</li></ul> |
| Trial conduct | <ul><li>Geographical dispersion of patients to cause difficulties of patient accrual</li><li>Lack of awareness by patients about rare disease to affect their attitude and willingness to participate in clinical trials</li><li>Insufficient engagement of healthcare system and local communities hampering recruitment of rare disease trials</li><li>Insufficient connection and partnership among various parties involved</li></ul> | <ul><li>Use of RWD for recruitment of diversified patients</li><li>Use of digital healthcare technologies to enhance participation of rare disease patients</li><li>Decentralized clinical trials for remote conduct of trial-related activities and monitoring</li></ul> |

Table 1 – continued from previous page

| Trial Aspects | Challenges | Strategies |
|---|---|---|
| Data analysis | • Use of external controls to introduce potential selection bias<br>• Heterogeneity of patient population to reduce study power<br>• Insufficient power for clinical benefits if surrogates are used as primary endpoints<br>• Uncertainty in BRA due to short follow-ups and single-arm design | • SATs with external controls<br>• Multiple endpoints, global tests, and repeated measures to overcome small sample size and improve efficiency<br>• Predictability of surrogates for clinical benefits<br>• Special considerations for BRA |
| Others | • Lack of enough market incentives<br>• Lack of high-quality RWD<br>• Difficulty to conduct pediatric trials<br>• Lack of coordinated effort | • Post-marketing studies to further demonstrate product effectiveness<br>• Expanded access program to collect more safety and effectiveness information<br>• Use of RWD and RWE in rare disease trials |

# 5  Concluding Remarks

There exists a wide scope of common challenges in rare disease drug development such as poor understanding of natural course of the disease, small populations, and diversified phenotypes and genotypes. This paper first reviews guidance documents and frameworks by regulatory agencies worldwide and then provides a landscape assessment from a statistical perspective on challenges in the design, conduct, and analysis of clinical trials of drugs treating rare diseases. While an overall development strategy and statistical considerations to address the aforementioned challenges are discussed, some novel design features (e.g., master protocols) and/or special implementation plan (e.g., DCTs with the use of DHTs) may be incorporated in rare disease trials, which may have some implications on the trial estimands; see [80] for precise construction of estimands in master protocols and [142] for considerations in defining an estimand in DCTs.

On the other hand, it is important to realize that developing drugs for rare diseases requires extraordinary effort and collaboration among researchers, clinicians, regulatory agencies, biopharmaceutical companies, patients and patient organizations. In particular, along the development pathway, it is critical for drug developers to (1) identify the patient needs, (2) develop, modify, and select fit-for-purpose clinical outcomes, (3) incorporate clinical outcome assessment into endpoints for regulatory decision-making, and (4) engage relevant regulatory agencies throughout the clinical development to ensure transparency, efficiency, and regulatory compliance.

## Abbreviations

| | |
|---|---|
| BRA | Benefit-risk assessment |
| COMP | Committee for Orphan Medicinal Products |
| DHTs | Digital healthcare technologies |
| DCTs | Decentralized clinical trials |
| EAP | Early access program |
| ECTs | Externally controlled trials |
| IP | Investigational product |
| MoA | Mechanism of action |
| ODA | Orphan Drug Act |
| PD | Pharmacodynamics |
| PK | Pharmacokinetics |
| PMC | Post-marketing commitment |
| PMR | Post-marketing requirement |
| PREA | Pediatric Research Equity Act |
| RCTs | Randomized controlled trials |
| RWD | Real-world data |
| RWE | Real-world evidence |
| SATs | Single-arm trials |

## Acknowledgements

The authors thank Dr. Mark Levenson of the FDA whose constructive comments have helped improve the presentation of the paper.

## Authors' contributions

## Funding

There is no funding support to this study.

## Data availability

No clinical data are involved in this research.

# Declaration

## Ethics approval and consent to participate

Not applicable.

## Consent for publication

Not applicable.

## Competing interest

The authors report there are no competing interests to declare. The FDA had no role in data collection, management, or analysis. The views expressed are those of the authors and not necessarily those of the US FDA.

# References

1. WHO. Rare diseases. World Health Organization (WHO): Geneva, Switzerland, 2021. Accessed on July 8, 2024.

2. FDA. Orphan Drug Act. Food and Drug Administration, U.S. Department of Health and Human Services, 1983. Retrieved 27 October 2015, accessed August 27, 2023.

3. Antoni Montserrat Moliner and Jaroslaw Waligora. The European Union policy in the field of rare diseases. *Rare Diseases Epidemiology: Update and Overview*, pages 561–587, 2017.

4. Lin Wang. Report on Definition and Research for Rare Diseases in China. Shanghai Foundation for Rare Diseases, 2021. Accessed on July 8, 2024.

5. Tomoe Uchida, Yoshimitsu Takahashi, Hiromitsu Yamashita, Yuriko Nakaoku, Tomoko Ohura, Takashi Okura, Yuko Masuzawa, Masayoshi Hosaka, Hiroshi Kobayashi, and Tami Sengoku. Evaluation of Clinical Practice Guidelines for Rare Diseases in Japan. *JMA Journal*, 5(4):460–470, 2022.

6. Ghada Mohammed Abozaid, Katie Kerr, Amy McKnight, and Hussain A Al-Omar. Criteria to define rare diseases and orphan drugs: a systematic review protocol. *BMJ Open*, 12(7):e062126, 2022.

7. NCATS. Rare Disease Day at NIH 2023. National Center for Advancing Translational Science, National Institutes of Health, U.S. Department of Health and Human Services, 2023. Accessed August 28, 2023.

8. Cameron Fox. Rare diseases: how can we improve diagnosis and treatment? World Economic Forum, 2023. Accessed on July 10, 2024.

9. TLD Endocrinology. Rare diseases: individually rare, collectively common. *The Lancet Diabetes & Endocrinology*, 11(3):139, 2023.

10. Shruti Marwaha, Joshua W Knowles, and Euan A Ashley. A guide for the diagnosis of rare and undiagnosed disease: beyond the exome. *Genome Medicine*, 14(1):1–22, 2022.

11. Lewis J Fermaglich and Kathleen L Miller. A comprehensive study of the rare diseases and conditions targeted by orphan drug designations and approvals over the forty years of the Orphan Drug Act. *Orphanet Journal of Rare Diseases*, 18(1):1–8, 2023.

12. NORD. Orphan Drugs in the United States: An Examination of Patents and Orphan Drug Exclusivity. National Organization for Rare Disorders, 2021.

13. Erika F Augustine, Heather R Adams, and Jonathan W Mink. Clinical trials in rare disease: challenges and opportunities. *Journal of Child Neurology*, 28(9):1142–1150, 2013.

14. Alice Cheng and Zhi Xie. Challenges in orphan drug development and regulatory policy in China. *Orphanet Journal of Rare Diseases*, 12(13):1–8, 2017.

15. Lucas Kempf, Jonathan C Goldsmith, and Robert Temple. Challenges of developing and conducting clinical trials in rare disorders. *American Journal of Medical Genetics Part A*, 176(4):773–783, 2018.

16. Diogo A Fonseca, Inês Amaral, Ana Catarina Pinto, and Maria Dulce Cotrim. Orphan drugs: major development challenges at the clinical stage. *Drug Discovery Today*, 24(3):867–872, 2019.

17. Erik Tambuyzer, Benjamin Vandendriessche, Christopher P Austin, Philip J Brooks, Kristina Larsson, Katherine I Miller Needleman, James Valentine, Kay Davies, Stephen C Groft, and Robert Preti. Therapies for rare diseases: therapeutic modalities, progress and challenges ahead. *Nature Reviews Drug Discovery*, 19(2):93–111, 2020.

18. The Lancet Global Health. The landscape for rare diseases in 2024, 2024.

19. Kathleen R Bogart and Veronica L Irvin. Health-related quality of life among adults with diverse rare disorders. *Orphanet Journal of Rare Diseases*, 12:1–9, 2017.

20. Annemieke Aartsma-Rus, Marc Dooms, and Yann Le Cam. Orphan medicine incentives: how to address the unmet needs of rare disease patients by optimizing the European orphan medicinal product landscape guiding principles and policy proposals by the European expert group for orphan drug incentives (OD Expert Group). *Frontiers in Pharmacology*, 12:3666, 2021.

21. Marilyn J Field and Thomas F Boat. *Rare diseases and orphan products: accelerating research and development*. National Academies Press, 2010.

22. Emilie Neez, Arianna Gentilini, and Adam Hutchings. Addressing unmet needs in extremely rare and paediatric-onset diseases: how the biopharmaceutical innovation model can help identify current issues and find potential solutions. The European Federation of Pharmaceutical Industry Associations (EFPIA), 2021. Accessed August 28, 2023.

23. FDA. Demonstrating Substantial Evidence of Effectiveness for Human Drug and Biological Products: Guidance for Industry, 2019.

24. FDA. *Guidance for Industry. Providing Clinical Evidence of Effectiveness for Human Drug and Biological Products*. US Department of Health and Human Services, 1998.

25. Jie Chen, Susan Gruber, Hana Lee, Shiowjen Lee, Haitao Chu, Haijun Tian, Yan Wang, Weili He, Thomas Jemielita, Susan Gruber, Yang Song, Roy Tamura, Lu Tian, Yihua Zhao, Yong Chen, Mark van der Laan, and Lei Nie. Rare Disease Drug Development: Use of Real-World Data and Real-World Evidence. *(In preparation)*, 2023.

26. FDA. Pediatric Rare Diseases–A Collaborative Approach for Drug Development Using Gaucher Disease as a Model–Guidance for Industry. Food and Drug Administration, U.S. Department of Health and Human Services, 2017.

27. FDA. Rare Diseases: Early Drug Development and the Role of Pre-IND Meetings–Draft Guidance for Industry. Food and Drug Administration, U.S. Department of Health and Human Services, 2018.

28. FDA. Slowly Progressive, Low-Prevalence Rare Diseases with Substrate Deposition That Results from Single Enzyme Defects: Providing Evidence of Effectiveness for Replacement or Corrective Therapies–Guidance for Industry. Food and Drug Administration, U.S. Department of Health and Human Services, 2018.

29. FDA. Rare Diseases: Common Issues in Drug Development (Draft Guidance). US Food and Drug Administration, Silver Spring, MD, 2019.

30. FDA. Rare Diseases: Natural History Studies for Drug Development. Guidance for Industry (Draft guidance). US Food and Drug Administration, Silver Spring, MD, 2019.

31. FDA. Rare Pediatric Disease Priority Review Vouchers–Guidance for Industry. Food and Drug Administration, U.S. Department of Health and Human Services, 2019.

32. FDA. Human Gene Therapy for Rare Diseases–Draft Guidance for Industry. Food and Drug Administration, U.S. Department of Health and Human Services, 2020.

33. FDA. Rare Diseases: COnsiderations for the Development of Drugs and Biological Products. The United States Food and Drug Administration, 2023.

34. FDA. Clinical Trial Considerations to Support Accelerated Approval of Oncology Therapeutics—Guidance for Industry. U.S. Department of Health and Human Services, Food and Drug Administration, 2023.

35. FDA. Considerations for the Design and Conduct of Externally Controlled Trials for Drug and Biological Products. U.S. Department of Health and Human Services, Food and Drug Administration, 2023.

36. FDA. Considerations for the Use of Real-World Data and Real-World Evidence To Support Regulatory Decision-Making for Drug and Biological Products. US Food and Drug Aministration, Department of Health and Human Services, 2023.

37. EMA. Committee for Orphan Medicinal Products (COMP). European Medicines Agency, 2019.

38. EMA. Guideline on Clinical Trials in Small Populations. Committee for Medicinal Products for Human Use (CHMP), European Medicines Agency, 2006.

39. EMA. Rare diseases, orphan medicines: Getting the facts straight. European Medicines Agency, 2018.

40. EMA. Points to consider on the estimation and reporting on the prevalence of a condition for the purpose of orphan designation. Committee for Orphan Medicinal Products (COMP), European Medicines Agency, 2019. Accessed September 2, 2023.

41. EMA. COMP work plan 2023. Committee for Orphan Medicinal Products (COMP), European Medicines Agency, 2023. Accessed September 2, 2023.

42. Yanqin Lu, Qingxia Gao, Xiuzhi Ren, Junfeng Li, Dan Yang, Zijian Zhang, and Jinxiang Han. Incidence and prevalence of 121 rare diseases in China: Current status and challenges: 2022 revision. *Intractable & Rare Diseases Research*, 11(3):96–104, 2022.

43. NMPA. Guidance on Clinical Development of Drugs for Rare Diseases. Center for Drug Evaluation, China National Medical Product Administration, 2022. Accessed: April 11, 2022.

44. NMPA. Statistical Guidance on Conducting Clinical Studies for Drug Development for Rare Diseases. Center for Drug Evaluation, China National Medical Products Administration, 2022. Accessed: March 10, 2023.

45. NMPA. Guidance on Natural History Studies of Rare Diseases for Drug Development. Center for Drug Evaluation, China National Medical Products Administration, 2023. Accessed: March 10, 2023.

46. MHLW. Overview of Orphan Drug/Medical Device Designation System. Ministry of Health, Labour and Welfare (MHLW) of Japan, 2023. Accessed: Semptember 2, 2023.

47. MHLW. Partial Revision of "Designation of Orphan Drugs etc.". Pharmaceutical Evaluation Division & Medical Device Evaluation Division, Pharmaceutical Safety Bureau, Ministry of Health, Labour and Welfare, 2024. Accessed by August 11, 2024.

48. MHLW. Questions and Answers (Q&A) for Designation of Orphan Drugs etc. Pharmaceutical Evaluation Division, Pharmaceutical Safety Bureau, Ministry of Health, Labour and Welfare, 2024. Accessed by August 11, 2024.

49. CORD. Rare Disease Day 2024 Summit. Canadian Organization for Rare Disorders, 2024.

50. CADTH. Framework for Drugs for Rare Diseases. Canadian Agency for Drugs and Technologies in Health, 2021.

51. CAPRA. Rare Diseases and Orphan Drugs Regulatory Framework in Canada: Recent Initiatives by Government of Canada. Canadian Association of Professionals in Regulatory Affairs, 2023. Accessed May 12, 2024.

52. MFDS. Regulation on Designation of Orphan Drugs. The Ministry of Food and Drug Safety (MFDS), South Korea, 2018. Accessed May 12, 2024.

53. Minseong Kim. Orphan Drug Designation Process in South Korea. Global Regulatory Partners, Inc., 2022. Accessed May 12, 2024.

54. UN. *World Population Ageing 2015*. Department of Economics and Social Affairs, United Nations, New York, 2015.

55. Roberto Giugliani, Silvia Castillo Taucher, Sylvia Hafez, Joao Bosco Oliveira, Mariana Rico-Restrepo, Paula Rozenfeld, Ignacio Zarante, and Claudia Gonzaga-Jauregui. Opportunities and challenges for newborn screening and early diagnosis of rare diseases in Latin America. *Frontiers in Genetics*, 13:1053559, 2022.

56. Daniel Wainstock and Amiel Katz. Advancing rare disease policy in Latin America: a call to action. *The Lancet Regional Health–Americas*, 18:1–7, 2023.

57. GlobalData. Over 25% of Rare Disease Trials Are Terminated Due to Low Patient Accrual Rates. GlobalData Pharma Intelligence Center, 2021. Accessed May 12, 2024.

58. CP Austin, CM Cutillo, LPL Lau, AH Jonker, A Rath, D Julkowska, D Thomson, SF Terry, B de Montleau, and D Ardigo. Future of rare diseases research 2017-2027: an IRDiRC perspective. *Clinical and Translational Science*, 11(1):21–27, 2018.

59. Dídac Casas-Alba, Janet Hoenicka, Alba Vilanova-Adell, Lourdes Vega-Hanna, and Jordi Pijuan. Diagnostic strategies in patients with undiagnosed and rare diseases. *Journal of Translational Genetics and Genomics*, 6:322–332, 2022.

60. Galliano Zanello, Chun-Hung Chan, and David A Pearce. Recommendations from the IRDiRC Working Group on methodologies to assess the impact of diagnoses and therapies on rare disease patients. *Orphanet Journal of Rare Diseases*, 17(1):1–10, 2022.

61. Takeya Adachi, Ayman W El-Hattab, Ritu Jain, Katya A Nogales Crespo, Camila I Quirland Lazo, Maurizio Scarpa, Marshall Summar, and Duangrurdee Wattanasirichaigoon. Enhancing Equitable Access to Rare Disease Diagnosis and Treatment around the World: A Review of Evidence, Policies, and Challenges. *International Journal of Environmental Research and Public Health*, 20(6):4732, 2023.

62. Joe TR Clarke, Doug Coyle, Gerald Evans, Janet Martin, and Eric Winquist. Toward a functional definition of a rare disease for regulatory authorities and funding agencies. *Value in Health*, 17(8):757–761, 2014.

63. FDA. FDA In Brief: FDA takes new steps to advance natural history studies for accelerating novel treatments for rare diseases. The U.S. Food and Drug Administration, 2019. Accessed: November 20, 2023.

64. Pamela Gavin. The importance of natural histories for rare diseases. *Expert Opinion on Orphan Drugs*, 3(8):855–857, 2015.

65. Thomas Ogorka and Gajendra Chanchu. The Impetus for Natural History Studies in Rare Disease R&D. *Applied Clinical Trials*, 27(4):18–21, 2018.

66. Jemima E Mellerio. The challenges of clinical trials in rare diseases. *British Journal of Dermatology*, 187(4):453–454, 2022.

67. Gerald F Cox. The art and science of choosing efficacy endpoints for rare disease clinical trials. *American Journal of Medical Genetics Part A*, 176(4):759–772, 2018.

68. Emil D Kakkis, Mary ODonovan, Gerald Cox, Mark Hayes, Federico Goodsaid, PK Tandon, Pat Furlong, Susan Boynton, Mladen Bozic, and May Orfali. Recommendations for the development of rare disease drugs using the accelerated approval pathway and for qualifying biomarkers as primary endpoints. *Orphanet Journal of Rare Diseases*, 10:1–17, 2015.

69. Wim Van der Elst, Geert Molenberghs, et al. *Surrogate Endpoints in Rare Diseases*, pages 257–74. Boca Raton: Chapman & Hall/CRC, 2016.

70. Carl Heneghan, Ben Goldacre, and Kamal R Mahtani. Why clinical trial outcomes fail to translate into benefits for patients. *Trials*, 18:1–7, 2017.

71. Rashmi Ashish Kadam, Sanghratna Umakant Borde, Sapna Amol Madas, Sundeep Santosh Salvi, and Sneha Saurabh Limaye. Challenges in recruitment and retention of clinical trial subjects. *Perspectives in Clinical Research*, 7(3):137–143, 2016.

72. James K Stoller. The challenge of rare diseases. *Chest*, 153(6):1309–1314, 2018.

73. IC Verma, A El-Beshlawy, A Tylki-Szymańska, A Martins, Y-L Duan, T Collin-Histed, M Schoneveld Van Der Linde, R Mansour, VC Dũng, and Pramod K Mistry. Transformative effect of a Humanitarian Program for individuals affected by rare diseases: Building support systems and creating local expertise. *Orphanet Journal of Rare Diseases*, 17(1):87, 2022.

74. Vivek Subbiah. The next generation of evidence-based medicine. *Nature Medicine*, 29(1):49–58, 2023.

75. Birutė Tumienė, Augutė Juozapavičiūtė, and Vytenis Andriukaitis. Rare diseases: still on the fringes of universal health coverage in Europe. *The Lancet Regional Health–Europe*, 37, 2024.

76. Jie Chen, Ying Lu, and Shivaani Kummar. Increasing Patient Participation in Oncology Clinical Trials. *Cancer Medicine*, 12(3):2219–2226, 2023.

77. Rima Nabbout, Galliano Zanello, Dixie Baker, Lora Black, Isabella Brambilla, Orion J Buske, Laurie S Conklin, Elin Haf Davies, Daria Julkowska, and Yeonju Kim. Towards the international interoperability of clinical research networks for rare diseases: recommendations from the IRDiRC Task Force. *Orphanet Journal of Rare Diseases*, 18(1):1–10, 2023.

78. Mahta Jahanshahi, Keith Gregg, Gillian Davis, Adora Ndu, Veronica Miller, Jerry Vockley, Cecile Ollivier, Tanja Franolic, and Sharon Sakai. The use of external controls in FDA regulatory decision making. *Therapeutic Innovation & Regulatory Science*, 55(5):1019–1035, 2021.

79. PS Mishra-Kalyani, L Amiri Kordestani, DR Rivera, H Singh, A Ibrahim, RA DeClaro, Y Shen, S Tang, R Sridhara, and PG Kluetz. External control arms in oncology: current use and future directions. *Annals of Oncology*, 2022.

80. Jie Chen, Sammy Yuan, Godwin Yung, Jingjing Ye, Hong Tian, and Jianchang Lin. Considerations for Master Protocols Using External Controls. *arXiv preprint arXiv:2307.05050*, 2023.

81. Ralf-Dieter Hilgers, Franz König, Geert Molenberghs, and Stephen Senn. Design and analysis of clinical trials for small rare disease populations. *Journal of Rare Diseases Research & Treatment*, 1(1):53–60, 2016.

82. Alisdair McNeill. Good genotype-phenotype relationships in rare disease are hard to find. *European Journal of Human Genetics*, 30(3):251–251, 2022.

83. Naji Bou Zeid and Victor Yazbeck. PI3k inhibitors in NHL and CLL: an unfulfilled promise. *Blood and Lymphatic Cancer: Targets and Therapy*, 13:1–12, 2023.

84. Nicole Gormley. Oncology Endpoint Development. US Food and Drug Administration Oncology Center of Excellence (OCE), 2024. Accessed by May 2, 2024.

85. Shein-Chung Chow and Wei Zhang. Statistical Evaluation of Clinical Trials Under COVID-19 Pandemic. *Therapeutic Innovation & Regulatory Science*, 54(6):1551–1556, 2020.

86. Mukund Nori. 'Negative' clinical trials in rare diseases and beyond: reclassification and potential solutions. *Future Rare Diseases*, 1(1):FRD5, 2021.

87. GK Raju, Karthik Gurumurthi, Reuben Domike, Dickran Kazandjian, Ola Landgren, Gideon M Blumenthal, Ann Farrell, Richard Pazdur, and Janet Woodcock. A benefit–risk analysis approach to capture regulatory decision-making: multiple myeloma. *Clinical Pharmacology & Therapeutics*, 103(1):67–76, 2018.

88. FDA. Benefit-Risk Assessment for New Drug and Biological Products. The US Food and Drug Administration, 2023.

89. FDA. Phosphatidylinositol 3-Kinase (PI3K) Inhibitors in Hematologic Malignancies. US Food and Drug Administration Oncologic Drugs Advisory Committee Meeting, 2024. Accessed by May 2, 2024.

90. NGF. Drug pricing, a complex issue affecting the rare disease community. National Gaucher Foundation, August 2018. Accessed: November 9, 2023.

91. EMA. Ophan Medicine Incentives—How to address the unmet needs of rare disease patients by transforming the European OMP landscape. European Expert Group on Orphan Drug Incentives, June 2021. Accessed: November 9, 2023.

92. Eric L Wan. Incentivizing Drug Development for Patients With Rare Diseases. Georgetown Medical Review, July 2023.

93. Zhiyao Zhao, Zhongyang Pei, Anxia Hu, Yuhui Zhang, and Jing Chen. Analysis of Incentive Policies and Initiatives on Orphan Drug Development in China: Challenges, Reforms and Implications. *Orphanet Journal of Rare Diseases*, 18(1):1–12, 2023.

94. Irin Tanaudommongkon, Shogo John Miyagi, Dionna J Green, Janelle M Burnham, John N van den Anker, Kyunghun Park, Johanna Wu, Susan K McCune, Lynne Yao, and Gilbert J Burckart. Combined pediatric and adult trials submitted to the US Food and Drug Administration 2012–2018. *Clinical Pharmacology & Therapeutics*, 108(5): 1018–1025, 2020.

95. Jack Cook, Dan Weiner, and J Robert Powell. Regarding combined pediatric and adult trials submitted to the US Food and Drug Administration 2012–2018. *Clinical Pharmacology & Therapeutics*, 109(5):1181, 2021.

96. Cesare Spadoni. Pediatric drug development: challenges and opportunities. *Current Therapeutic Research, Clinical and Experimental*, 90:119, 2019.

97. Hanns Lochmüller, Josep Torrent i Farnell, Yann Le Cam, Anneliene H Jonker, Lilian PL Lau, Gareth Baynam, Petra Kaufmann, Hugh JS Dawkins, Paul Lasko, and Christopher P Austin. The international rare diseases research consortium: policies and guidelines to maximize impact. *European Journal of Human Genetics*, 25(12): 1293–1302, 2017.

98. G Genes. Rare diseases, common challenges. *Nature Genetics*, 54:215, 2022.

99. Meghan C Halley, Hadley Stevens Smith, Euan A Ashley, Aaron J Goldenberg, and Holly K Tabor. A call for an integrated approach to improve efficiency, equity and sustainability in rare disease research in the United States. *Nature Genetics*, 54(3): 219–222, 2022.

100. Amy M Patterson, Megan OBoyle, Grace E VanNoy, and Kira A Dies. Emerging roles and opportunities for rare disease patient advocacy groups. *Therapeutic Advances in Rare Disease*, 4:1–18, 2023.

101. US Government. Code of federal regulations title 21. The Office of the Federal Register, National Archives and Records Administration, 2017. Accessed: November 13, 2023.

102. FDA. Demonstrating Substantial Evidence of Effectiveness With One Adequate and Well-Controlled Clinical Investigation and Confirmatory Evidence. U.S. Department of Health and Human Services Food and Drug Administration, 2023.
36

103. FDA. Adaptive designs for clinical trials of drugs and biologics – guidance for industry. US Department of Health and Human Services, Food and Drug Administration, 2019.

104. Sidi Wang, Kelley M Kidwell, and Satrajit Roychoudhury. Dynamic enrichment of Bayesian small-sample, sequential, multiple assignment randomized trial design using natural history data: a case study from Duchenne muscular dystrophy. *Biometrics*.

105. Heng Xu, Yi Liu, and Robert A Beckman. Adaptive endpoints selection with application in rare disease. *Statistics in Biopharmaceutical Research*, pages 1–9, 2023.

106. FDA. Enrichment Strategies for Clinical Trials to Support Determination of Effectiveness of Human Drugs and Biological Products: Guidance for Industry. US Food and Drug Administration, Silver Spring, MD, 2019.

107. Chiara Pizzamiglio, Hilary J Vernon, Michael G Hanna, and Robert DS Pitceathly. Designing clinical trials for rare diseases: unique challenges and opportunities. *Nature Reviews Methods Primers*, 2(1):13, 2022.

108. FDA. Master Protocols: Efficient Clinical Trial Design Strategies to Expedite Development of Oncology Drugs and Biologics. Guidance for Industry. US Food and Drug Administration, 2022.

109. Joan M Korth-Bradley. Regulatory framework for drug development in rare diseases. *The Journal of Clinical Pharmacology*, 62:S15–S26, 2022.

110. Ruo-Jing Li, Lian Ma, Fang Li, Liang Li, Youwei Bi, Ye Yuan, Yangbing Li, Yuan Xu, Xinyuan Zhang, and Jiang Liu. Model-Informed Approach Supporting Drug Development and Regulatory Evaluation for Rare Diseases. *The Journal of Clinical Pharmacology*, 62:S27–S37, 2022.

111. FDA. BEST (Biomarkers, EndpointS, and other Tools) Resource. FDA-NIH Biomarker Working Group, Food and Drug Administration (US), Silver Spring, MD, 2016. Date of access: Sept. 15, 2021.

112. S Amur, L LaVange, I Zineh, S Buckman-Garner, and J Woodcock. Biomarker qualification: toward a multiple stakeholder framework for biomarker development, regula-tory acceptance, and utilization. *Clinical Pharmacology & Therapeutics*, 98(1):34–46, 2015.

113. FDA. Rare Disease Endpoint Advancement Pilot Program. U.S. Department of Health and Human Services Food and Drug Administration, 2022. Accessed: November 13, 2023.

114. ICH. Choice of Control Group and Related Issues in Clinical Trials (E10) . International Conference on Harmonisation of Technical Requirements for Registration of Pharmaceuticals for Human Use, 2001.

115. Meng H Tan, Matthew Thomas, and Mark P MacEachern. Using registries to recruit subjects for clinical trials. *Contemporary Clinical Trials*, 41:31–38, 2015.

116. FDA. Digital Health Technologies for Remote Data Acquisition in Clinical Investigations. The US Food and Drug Aministration, Department of Health and Human Services, 2023.

117. FDA. Decentralized Clinical Trials for Drugs, Biological Products, and Devices. Food and Drug Administration, U.S. Department of Health and Human Services, 2023.

118. Mercedeh Ghadessi, Junrui Di, Chenkun Wang, Kiichiro Toyoizumi, Nan Shao, Chaoqun Mei, Charmaine Demanuele, Rui Tang, Gianna McMillan, and Robert A Beckman. Decentralized clinical trials and rare diseases: a Drug Information Association Innovative Design Scientific Working Group (DIA-IDSWG) perspective. *Orphanet Journal of Rare Diseases*, 18(1):79, 2023.

119. Shein-Chung Chow. *Innovative Statistics in Regulatory Science.* CRC Press, 2019.

120. Bo Yang, Yang Song, and Yijie Zhou, editors. *Drug Development for Rare Diseases*. Chapman & Hall, 2023.

121. Johan Verbeeck, Maya Dirani, Johann W Bauer, Ralf-Dieter Hilgers, Geert Molenberghs, and Rima Nabbout. Composite endpoints, including patient reported outcomes, in rare diseases. *Orphanet Journal of Rare Diseases*, 18(1):262, 2023.

122. Stuart J Pocock, Cono A Ariti, Timothy J Collier, and Duolao Wang. The win ratio: a new approach to the analysis of composite endpoints in clinical trials based on clinical priorities. *European Heart Journal*, 33(2):176–182, 2012.

123. Scott R Evans, Daniel Rubin, Dean Follmann, Gene Pennello, W Charles Huskins, John H Powers, David Schoenfeld, Christy Chuang-Stein, Sara E Cosgrove, Vance G Fowler Jr, Ebbing Lautenback, and Henry F Chambers. Desirability of outcome ranking (DOOR) and response adjusted for duration of antibiotic risk (RADAR). *Clinical Infectious Diseases*, 61(5):800–806, 2015.

124. Robin Ristl, Susanne Urach, Gerd Rosenkranz, and Martin Posch. Methods for the analysis of multiple endpoints in small populations: a review. *Journal of Biopharmaceutical Statistics*, 29(1):1–29, 2019.

125. M Morris, W Lee, and Y Wang. Evaluation of Testing Methods for Multiple Endpoint in Small Sized Trials: Application to Rare Diseases. CDER ORISE research project, the US Food and Drug Administration, 2019.

126. Rima Izem and Robert McCarter. Randomized and non-randomized designs for causal inference with longitudinal data in rare disorders. *Orphanet Journal of Rare Diseases*, 16:1–9, 2021.

127. Tze L Lai and Philip W Lavori. Innovative clinical trial designs: toward a 21st-century health care system. *Statistics in Biosciences*, 3(2):145–168, 2011.

128. Tze Leung Lai, Philip W Lavori, Mei-Chiung I Shih, and Branimir I Sikic. Clinical trial designs for testing biomarker-based personalized therapies. *Clinical Trials*, 9(2): 141–154, 2012.

129. Cong Chen, Keaven Anderson, Devan V Mehrotra, Eric H Rubin, and Archie Tse. A 2-in-1 adaptive phase 2/3 design for expedited oncology drug development. *Contemporary Clinical Trials*, 64:238–242, 2018.

130. Cong Chen and Eric H Rubin. Adaptive phase 2/3 designs for oncology drug development–Time to hedge. *Contemporary Clinical Trials*, 125:107047, 2023.

131. IMI-PROTECT. Recommendations for the methodology and visualisation techniques to be used in the assessment of benefit and risk of medicines. The PROTECT consortium (Pharmacoepidemiological Research on Outcomes of Therapeutics by a European ConsorTium, www.imiprotect.eu), 2023.

132. ESMO. The ESMO-Magnitude of Clinical Benefit Scale (ESMO-MCBS). European Society for Medical Oncology, 2023.

133. Richard Huckle. Challenges in benefit-risk assessment of orphan drugs. *Regulatory Rapport*, 12(2):9, 2015. Accessed on July 10, 2024.

134. T Morel, S Aymé, D Cassiman, S Simoens, M Morgan, and M Vandebroek. Quantifying benefit-risk preferences for new medicines in rare disease patients and caregivers. *Orphanet Journal of Rare Diseases*, 11:1–12, 2016.

135. FDA. Guidance for Industry: Expedited Programs for Serious Conditions – Drugs and Biologics. 2014.

136. FDA. *Guidance for Industry: Postmarketing Studies and Clinical Trials–Implementation of Section 505 (o)(3) of the Federal Food, Drug, and Cosmetic Act*. US Food and Drug Administration, 2012.

137. William C Maier, Ronald A Christensen, and Patricia Anderson. Post-approval studies for rare disease treatments and orphan drugs. In Domenica de la Paz, Manuel Posadaand Taruscio and Stephen C Groft, editors, *Rare Diseases Epidemiology: Update and Overview*, pages 197–205. Springer, 2017.

138. FDA. Postmarketing Studies and Clinical Trials: Determining Good Cause for Non-compliance with Section 505(o)(3)(E)(ii) of the Federal Food, Drug, and Cosmetic Act Guidance for Industry. U.S. Department of Health and Human Services Food and Drug Administration, 2023.

139. FDA. Expanded Access. The US Food and Drug Administration, 2024. Accessed May 13, 2024.

140. Tobias B Polak, David GJ Cucchi, Joost van Rosmalen, Carin A Uyl-de Groot, and Jonathan J Darrow. Generating evidence from expanded access use of rare disease medicines: challenges and recommendations. *Frontiers in Pharmacology*, 23(13):1917, 2022.

141. FDA. Framework for FDA's Real-World Evidence Program. U.S. Department of Health and Human Services, Food and Drug Administration, Silver Spring, Maryland, USA, 2018.

142. ASA RWE SWG. Decentralized Clinical Trials in the Era of Real-World Evidence: A Statistical Perspective. *(Submitted)*, 2024.